\documentclass{emulateapj}
\usepackage{amsmath}
\usepackage{amssymb}
\usepackage{latexsym}
\usepackage{textcomp}
\usepackage{graphicx}
\usepackage{subfigure}
\usepackage{lscape}
\usepackage{longtable}
\usepackage{apjfonts}

\slugcomment{Accepted to ApJ, July 26, 2008}

\begin{document}
\title{Theoretical Radii of Extrasolar Giant Planets: \\
the Cases of TrES-4, XO-3b, and HAT-P-1b}

\shorttitle{THEORETICAL RADII OF EXTRASOLAR GIANT PLANETS}
\shortauthors{LIU ET AL.}
\author{\sc Xin Liu\altaffilmark{1}, Adam Burrows\altaffilmark{1,2}, and Laurent Ibgui\altaffilmark{2}}

\altaffiltext{1}{Department of Astrophysical Sciences, Princeton
University, Peyton Hall -- Ivy Lane, Princeton, NJ 08544;
xinliu@astro.princeton.edu, burrows@astro.princeton.edu}

\altaffiltext{2}{Department of Astronomy and Steward Observatory,
                 The University of Arizona, Tucson, AZ \ 85721; laurent@as.arizona.edu}

\begin{abstract}
To explain their observed radii, we present theoretical radius-age
trajectories for the extrasolar giant planets (EGPs) TrES-4,
XO-3b, and HAT-P-1b.  We factor in variations in atmospheric
opacity, the presence of an inner heavy-element core, and possible
heating due to orbital tidal dissipation. A small, yet non-zero,
degree of core heating is needed to explain the observed radius of
TrES-4, unless its atmospheric opacity is significantly larger
than a value equivalent to that at 10$\times$solar metallicity
with equilibrium molecular abundances. This heating rate is
reasonable, and corresponds for an energy dissipation parameter
($Q_p$) of $\sim$10$^{3.8}$ to an eccentricity of $\sim$0.01,
assuming 3$\times$solar atmospheric opacity and a heavy-element
core of $M_c = 30$ $M_{\oplus}$. For XO-3b, which has an observed
orbital eccentricity of $0.26$, we show that tidal heating needs
to be taken into account to explain its observed radius.
Furthermore, we reexamine the core mass needed for HAT-P-1b in
light of new measurements and find that it now generally follows
the correlation between stellar metallicity and core mass
suggested recently. Given various core heating rates, theoretical
grids and fitting formulae for a giant planet's equilibrium radius
and equilibration timescale are provided for planet masses $M_p=$
0.5, 1.0, and 1.5 $M_J$ with $a =$ 0.02-0.06 AU, orbiting a G2V
star. When the equilibration timescale is much shorter than that
of tidal heating variation, the ``effective age'' of the planet is
shortened, resulting in evolutionary trajectories more like those
of younger EGPs. Motivated by the work of
\citet{jackson08a,jackson08b}, we suggest that this
effect could indeed be important in better
explaining some observed transit radii.
\end{abstract}

\keywords{planetary systems --- planets and satellites: general
--- stars: individual (HAT-P-1, GSC 02620-00648, XO-3)}

%------------------------------------------------------------------------------
\section{Introduction}\label{sec:intro}

As of the writing of this paper, an astounding 47 transiting
extrasolar giant planets (EGPs) have been discovered\footnote{See
J. Schneider's Extrasolar Planet Encyclopaedia at
http://exoplanet.eu, the Geneva Search Programme at
http://exoplanets.eu, and the Carnegie/California compilation at
http://exoplanets.org}. For transiting planets, the
inclination/planet-mass degeneracy is resolved and the photometric
dip in the stellar flux during the transit yields the planet's
radius ($R_p$). Theory then attempts to explain the measured radii
\citep{guillot96,burrows00,burrows03,burrows04b,burrows07b,bodenheimer01,
bodenheimer03,baraffe03,baraffe04,baraffe08,chabrier04,fortney07a,laughlin05}.
Importantly, the comparison between theory and measurement must be
done for a given stellar type, orbital distance, planet
mass, and age. The latter is poorly measured, but crucially
important (see, e.g. \S\ref{sec:sum} re HD 209458b).

\citet{burrows07b} modeled the theoretical evolution of the radii
of the 14 transiting EGPs known at the time. These authors suggest
that there are two radius anomalies in the transiting EGP family,
of which the smaller-radius anomaly is a result of the presence of
dense cores \citep{mizuno80,pollack96,hubickyj04}, whereas the
larger-radius anomaly might be explained by the enhanced
atmospheric opacities which slow down the heat loss of the core.
They also discussed the effects on planet structure of the
possible extra heat source in the interior, yet found no obvious
correlation between the requisite power and the intercepted
stellar power. Note that none of the transiting EGPs modeled by
\citet{burrows07b} is known to have a highly eccentric orbit.
However, recently several transiting EGPs with significantly
non-zero eccentricities ($e \gtrsim 0.15$) have been discovered,
including the EGPs XO-3b \citep{johnskrull07}, HAT-P-2b
\citep[a.k.a. HD 147506b;][]{bakos07b,loeillet07}, HD 17156b
\citep{barbieri07,gillon07d,narita07,irwin08}, and the ``hot
Neptune'' GJ 436b \citep{gillon07c,deming07a,demory07}.

At least for those systems with highly eccentric orbits, heating
due to orbital tidal dissipation
\citep{bodenheimer01,bodenheimer03} should be incorporated into
theoretical models for the radius-age trajectories.  A preliminary
exploration of this in a restricted context motivates the present
paper.  When the theoretical radius evolution calculations are
tailored to a system's specific planet mass, age, primary stellar
properties, and orbital distance (as they must), the current radii
of most of the known transiting EGPs can be explained by the
theoretical radius models of \citet{burrows07b}. In many
instances, a higher-density core mass provides an even better fit,
and extra internal heat sources are not required. However, at
least three transiting EGPs seem to be exceptional in some way.
The planets TrES-4 and XO-3b (for its large-radius solution based
on stellar parameters from spectral synthesis modeling) are cited
by their discoverers \citep{mandushev07,johnskrull07} as
anomalously large and inconsistent with extant theoretical models.
In addition, \citet{burrows07b} found that HAT-P-1b deviated from
the core-mass stellar-metallicity relationship followed by many of
the transiting EGPs that they modeled \citep[for the core-mass
stellar-metallicity correlation, also see][]{guillot06}.

Therefore, with this paper, we focus on this small subset of three
interesting objects to determine how their radii can indeed be
explained with minimal assumptions, that nevertheless can include
tidal heating. We find that tidal heating in the core, given
measured (XO-3b) and possibly non-zero (TrES-4 and HAT-P-1b)
eccentricities, which nevertheless are still consistent with the
upper limits, can naturally explain the measured radii. For
HAT-P-1b with its new age estimate, we find that a core of
reasonable size can now be accommodated.  Importantly, when tidal
heating needs to be invoked, a canonical tidal dissipation
parameter, ``$Q_p$''\citep{goldreich66}\footnote{Note that the
$Q_p$ parameter used throughout this paper corresponds to the $Q_p'$
in \citet{mardling07}.}, with a value near 10$^{4-6}$, along with
the measured eccentricity (or reasonable values consistent with
its current bounds), suffices to explain the measurements. Hence,
only simple extensions of the default evolution models that
incorporate a known process with canonical parameters are
required.  We postulate that all measured transiting EGP radii can
be explained by available theory when proper account is taken of
the measured planet-star system parameters, reasonable core masses
that follow the relationship between core mass and stellar
metallicity \citep{burrows07b,guillot06}, and tidal heating at the
expected rate for non-zero, but measured, eccentricities.
Exceptions to this might arise if it is determined that tidal
dissipation occurs predominantly in the atmosphere, not the
convective core, and/or if the eccentricity and semi-major axis
history must be factored into the tidal heating history of the
planet.  The latter effect is intriguing and has been suggested by
\citet{jackson08a,jackson08b}.

We describe our computational techniques and model assumptions in
\S\ref{sec:tech}. In \S\ref{sec:obs}, we review the measurements
of these three EGP systems, identify the discrepancies between the
observed planetary radii and those predicted by previous
theoretical models and present new theoretical radius-age
trajectories for them using tailored atmospheric boundary
conditions.  These new trajectories and theoretical radii include
the effects of tidal heating in the convective core for measured
or reasonable values of the orbital eccentricities and for a range
of values for the tidal parameter, $Q_p$.  In \S\ref{sec:heating},
we provide theoretical grids and fitting models for the
equilibrium planetary radius and the equilibration timescale,
given a certain set of planet mass, orbital distance, and tidal
heating rate for a G2V primary star. Finally, in \S\ref{sec:sum}
we summarize our results for each system, discuss the relevant
constraints obtained on their structural properties, and list
caveats concerning our model assumptions.

%------------------------------------------------------------------------------
\section{Computational Techniques and Model Assumptions}\label{sec:tech}

A detailed discussion of our computational techniques can be found
in \citet{burrows03} and \citet{burrows07b}. Here, we present only
a brief summary, along with our model assumptions. We generate
realistic atmospheres customized for the three EGPs, their
time-averaged orbital separations, and primary stars. The adopted
atmospheric boundary conditions incorporate irradiation using the
observed stellar luminosity and spectrum, and the measured planet
orbital distance. For planets with eccentric orbits, the
time-averaged insolation flux is employed in constructing the
atmospheric boundary conditions. The theoretical stellar spectra
of \citet{kuruz94} are adopted. For the given stellar spectrum and
flux (inferred from the luminosity and the planet orbital
distance), an $S$-$T_{eff}$-$g$ grid is calculated for the core
entropy, $S$, effective temperature, $T_{eff}$, and gravity, $g$,
using the discontinuous finite element (DFE) variant of the
spectral code TLUSTY \citep{hubeny95}. It is assumed that both the
stellar spectrum and flux are constant during the evolution.

We employ the Henyey evolutionary code of \citet{burrows97} for
the radius-age evolutionary calculations, using the function
$T_{eff}(S, g)$ for the interior flux, inverted from the table of
$S$, $T_{eff}$, and $g$, referred to above. The helium fraction
$Y_{\rm He}$ is assumed to be 0.25. We calculate models with
different atmospheric opacities, the effect of which can be
conveniently mimicked by using 1$\times$solar, 3$\times$solar, and
10$\times$solar abundance atmospheres. Note that the increase in
atmospheric opacity does not need to, and should not be due solely
to, increased metallicity. The effects of increased atmospheric
opacity and increased envelope heavy-element abundances are
decoupled, so that the implied increases in the heavy-element
burden of the envelope, if any, will not cancel the expansion
effect of enhanced atmospheric opacity \citep{burrows07b}. To
model the presence of a heavy-element core, a compressible ball of
olivine is placed in the center of the model planet, and pressure
continuity between the heavy-element core and the gaseous envelope
is ensured throughout the evolution. We adopt the \citet{saumon95}
equation of state (EOS) for the H$_2$/He envelope and the ANEOS by
\citet{thompson72} for olivine.

As noted, to model the atmospheric opacity, we use supersolar
metallicities (e.g., $3\times$solar, $10\times$solar) to mimic the
expansion effects of enhanced atmospheric opacity
\citep{burrows07b}. Possible causes for such enhanced opacities
might be supersolar metallicities in the atmosphere,
nonequilibrium chemistry, errors in the default opacities, and
thick hazes or absorbing clouds. Note that the
expansion effects of enhanced atmospheric opacity and the
shrinkage effects of increased envelope metallicities conceptually
decoupled in our models and that an increase of the envelope
heavy-element burden, will not necessarily cancel the expansion
effect due to enhanced atmospheric opacity. The effect of a
central heavy-element core on the planet radius is to shrink it
monotonically with core mass.

Given a non-zero orbital eccentricity, the tidal dissipation rate
is calculated using the formulation summarized in the Appendix
\citep{bodenheimer01,bodenheimer03,gu04}. We assume that the
planet is in synchronous rotation and that all the tidal heating
is in the convective core. Note that the effects of other core
energy dissipation mechanisms on planet structural evolution are
also implicitly addressed. As indicated in the Appendix, the tidal
heating rate is proportional to $f(e)/Q_p$, where $f(e) = e^2$
when $e \ll 1$. Also, the values of $Q_p$ for EGPs with masses
$M_p\sim$ $M_J$, although very uncertain, are estimated to be
$\sim10^4-10^7$ \citep{adams06,gu03,jackson08a,jackson08b}.
Therefore, we calculate typical heating rates using the
combination $e^2/Q_p$ for TrES-4 and HAT-P-1b, of which the
orbital eccentricities have been estimated to be $\ll 1$, if
nonzero at all. Models without external heat sources ($e^2/Q_p =
0$) are also presented.

%------------------------------------------------------------------------------
\section{Observed Properties and Theoretical Planetary Radii}\label{sec:obs}

Table \ref{tab:planet} displays the relevant observed quantities
and the corresponding references for the EGPs TrES-4, XO-3b, and
HAT-P-1b and their parent stars. These properties include
semi-major axis ($a$), orbital period and eccentricity ($e$),
stellar mass ($M_{\ast}$), radius ($R_{\ast}$), effective
temperature $T_{eff}$, metallicity ([Fe/H]$_{\ast}$), age, and
planetary mass ($M_p$) and radius ($R_p$). Also shown are the
stellar flux at the planet's substellar point, $F_p$, in units of
$10^9$ erg cm$^{-2}$ s$^{-1}$, and the ratio between the possible
tidal energy dissipation rate within the planet and the insolation
rate $\dot{E}_{{\rm tide}}/\dot{E}_{{\rm insolation}}$ in the unit
of $(Q_p/10^5)^{-1}$. $Q_p^{-1} \equiv \frac{1}{2\pi E_0}\oint
(-\frac{dE}{dt})dt$ is the specific dissipation function of the
planet, where $E_0$ is the maximum energy stored in the tidal
distortion and $-\frac{dE}{dt}$ is the rate of dissipation
\citep{goldreich66}. The parameters for TrES-4 and HAT-P-1b are
drawn from \citet{torres08}. These authors provide a uniform
analysis of transit light curves and stellar parameters based on
stellar evolution models, and a critical examination of the
corresponding errors. Since XO-3b has not been studied by
\citet{torres08}, we generate models using parameters
reported by \citet{johnskrull07} and \citet{winn08}.

Theoretical evolutionary trajectories are presented for TrES-4,
XO-3b, and HAT-P-1b under various assumptions about the
atmospheric opacity, the presence of a heavy-element core and
possible tidal heating. The transit radius effect
\citep{burrows07b} is also included in the models.

\subsection{TrES-4}

TrES-4 is the current record-holder for the lowest EGP density
\citep{mandushev07,torres08}. In the discovery work,
\citet{mandushev07} carried out spectroscopic observations with
the CfA Digital Speedometer \citep{latham92} , radial velocity
(RV) measurements with Keck, and transit photometry in the $z$
band with KeplerCam at the F. L. Whipple Observatory (FLWO) and in
the $B$ band using NASACam on the 0.8-m telescope at the Lowell
Observatory. Assuming [Fe/H]$_{\ast}$ = $0.0_{-0.2}^{+0.2}$ dex,
Mandushev et al. derived $T_{eff} = 6100_{-150}^{+150}$ K, $M_{\ast}
= 1.22_{-0.17}^{+0.17}$ $M_{\odot}$, $R_{\ast} =
1.738_{-0.092}^{+0.092}$ $R_{\odot}$, and an age of
$4.7_{-2.0}^{+2.0}$ Gyr for the star, and $M_p =
0.84_{-0.10}^{+0.10}$ $M_J$ and $R_p = 1.674_{-0.094}^{+0.094}$
$R_J$ for the planet. The orbital eccentricity was assumed to be
exactly zero in the fit. \citet{mandushev07} suggested that its
observed radius is too large to be explained by the theoretical
EGP models of \citet{burrows07b}, given its estimated mass, age
and insolation, even when the effects of higher atmospheric
opacities and the transit radius correction are considered.

The parameters of TrES-4 listed in Table \ref{tab:planet} are from
\citet{torres08}. These authors derived these parameters using the
RV measurements and transit photometry from \citet{mandushev07},
and the stellar atmospheric properties from Sozzetti {\it et al.}
(2008; in preparation). Note that their estimated planetary radius
$R_p = 1.751_{-0.062}^{+0.064}$ $R_J$ is $\sim$1-$\sigma$ larger
than the value of $R_p = 1.674_{-0.094}^{+0.094}$ $R_J$ of
\citet{mandushev07}.

\subsubsection{Results for TrES-4}\label{sec:res:tres4}

\begin{figure}
 \centering
    \includegraphics[width=89mm]{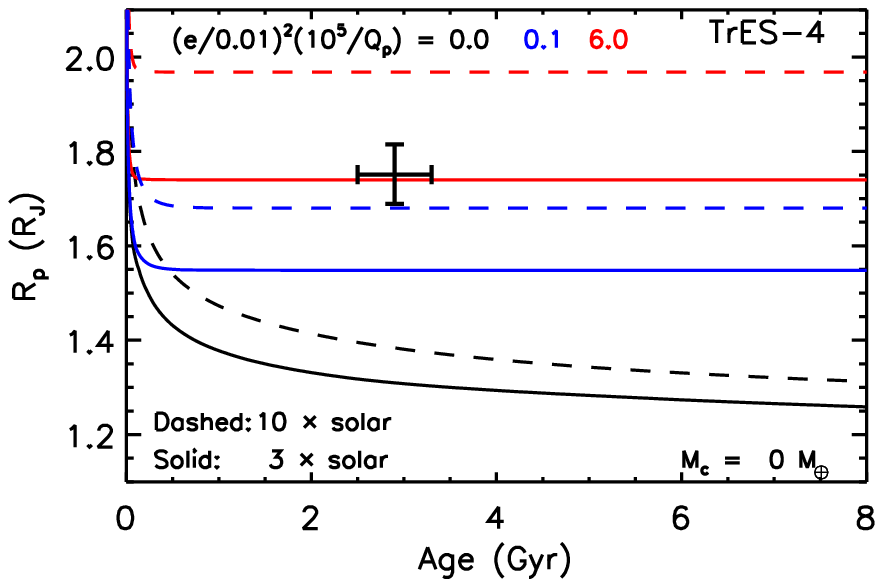}
    \includegraphics[width=89mm]{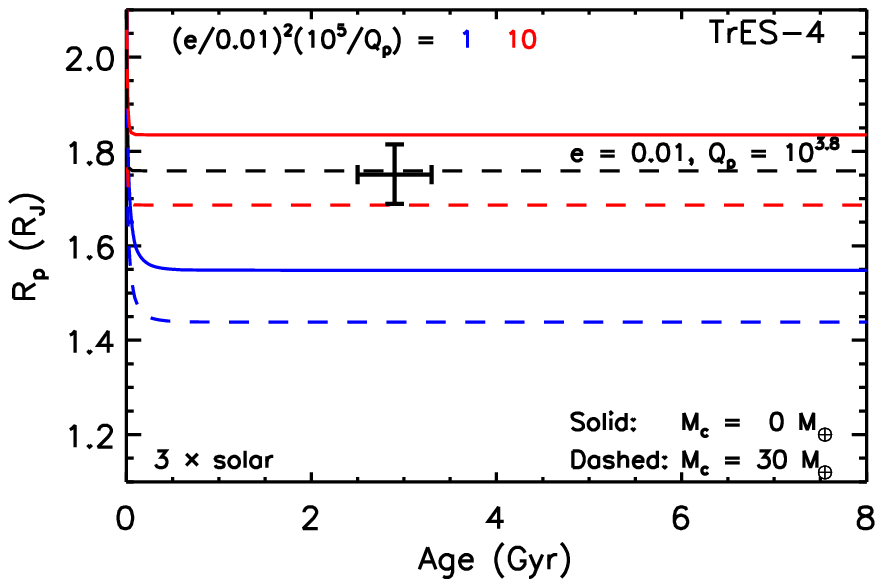}
   \caption{Theoretical planet radius $R_p$ ($R_J$) versus age (Gyr)
   for TrES-4. Also shown on both panels with error bars are the observed radius and
   age from \citet{torres08}. Various values are assumed concerning
   the atmospheric opacity, the presence of a heavy-element core, and the core heating due
   to tidal dissipation. Different colors correspond to different tidal heating rates,
   which are proportional to $e^2/Q_p$ when $e \ll 1$ (see the Appendix for more details).
   {\it Top}: This panel demonstrates the effect of enhanced atmospheric opacity under various
   heating powers.  Models assuming 3$\times$solar ($10\times$solar) atmospheric
   opacities are plotted as solid (dashed) curves, whereas the black, blue, and red curves
   correspond to $(e/0.01)^2(10^5/Q_p) = 0.0$, 0.1, and 6.0, all without a heavy-element core
   ($M_c$ = 0). {\it Bottom}: The effect of the presence of a heavy-element core is illustrated, where the
   dashed (solid) curves denote models with $M_c$ = 30 (0) $M_{\oplus}$. If TrES-4 follows the
   core-mass stellar-metallicity relation found by \citet{burrows07b}, then it should contain
   a heavy-element core with $M_c\sim$$20-40$ $M_{\oplus}$, given its stellar metallicity
   [Fe/H]$_{\ast}$ ($+0.14_{-0.09}^{+0.09}$ dex). Assuming 3$\times$solar atmospheric opacity,
   the model with $Q_p = 10^{3.8}$, $e \sim$0.01, and a heavy-element core of $M_c = 30$ $M_{\oplus}$
   ({\it black-dashed} curve) explains the observed radius well.}
   \label{fig:tres4}
\end{figure}
%-------------

We calculate the radius-age trajectories for TrES-4 using the
parameters from \cite{torres08}, taking into account the possible
effects of enhanced atmospheric opacities, and the presence of
tidal dissipation given a small, yet non-zero, orbital
eccentricity. Models with the presence of a heavy-element core are
also calculated. The value of $\dot{E}_{{\rm tide}}/\dot{E}_{{\rm insolation}}$
quoted for TrES-4 in Table \ref{tab:planet} is calculated assuming
$e = 0.01$, but using all the other parameters from
\cite{torres08} in which a circular orbit is assumed. More and
better transit observations are needed to better constrain its
true orbital eccentricity.

The first panel of Fig. \ref{fig:tres4} shows the theoretical
radii $R_p$ (in units of $R_J$) as a function of age (in units of Gyr) for
TrES-4, under various assumptions concerning the atmospheric opacity and the
level of tidal dissipation, without a heavy-element core
($M_c = 0$). Models assuming $3\times$ ($10\times$) solar
atmospheric opacities are shown as solid (dotted) curves. The
black curves show the models without any heat sources, whereas the
blue (red) curves depict those with a tidal heating rate assuming
$(e/0.01)^2(10^5/Q_p) = 0.1$ ($6.0$). Also shown with error bars
are the observed radius and age from \citet{torres08} (Table
\ref{tab:planet}).

It can be seen from Fig. \ref{fig:tres4} that if the radius and
age estimates of \citet{torres08} do not deviate much from their
true values (within the uncertainties), our models without any
heat sources produce radii which are $\sim$3 $\sigma$ too small.
Assuming the \citet{mandushev07} parameters, the theoretical radii
are still $\sim$2 $\sigma$ too small. The discrepancy will become
smaller for models with even higher atmospheric opacities. So it
is concluded that either the atmospheric opacity of TrES-4 is
unusually large (much higher than the equivalent of a
$10\times$solar metallicity, equilibrium mixture), or there are
extra heat sources in the core. The required heating power is very
modest; the model with $(e/0.01)^2(10^5/Q_p) = 0.1$ and
10$\times$solar atmospheric opacity produces theoretical radii
consistent with the 1-$\sigma$ lower bound of $R_p$ from
\citet{torres08}.

The models shown in the first panel of Fig. \ref{fig:tres4} do not
include any heavy-element cores. If TrES-4 follows the core-mass
stellar-metallicity relation studied by \citet{burrows07b}, then
there should be a heavy element core with $M_c\sim$$20-40$
$M_{\oplus}$, given its stellar metallicity [Fe/H]$_{\ast}$
($+0.14_{-0.09}^{+0.09}$ dex). The presence of such a
heavy-element core will shrink the planetary radius, and,
therefore, would require a higher tidal heating rate than do
models without a core to explain the observed radius. This effect
of including a heavy-element core in the models for TrES-4 is
shown in the second panel of Fig. \ref{fig:tres4}. Assuming
3$\times$solar atmospheric opacity, the model with $Q_p =
10^{3.8}$, $e \sim$0.01, and a heavy-element core of $M_c = 30$
$M_{\oplus}$ ({\it black-dashed}) explains the observed radius
well. Within 1 $\sigma$ uncertainties, the model with $Q_p =
10^{4.0}$, $e \sim$0.01, and a heavy-element core of $M_c = 30$
$M_{\oplus}$ ({\it red-dashed} curve) can also fit the observed
radius.

In summary, unless the atmospheric opacity of TrES-4 is unusually
large, core heating is required to explain its observed radius.
However, the required heating power is modest. A non-core model
with $(e/0.01)^2(10^5/Q_p) = 0.1$ produces radii consistent with
the 1-$\sigma$ lower boundary of $R_p$ from \citet{torres08},
assuming $10 \times$solar atmospheric opacity. The required energy
dissipation rates become larger for models with a heavy-element
core, but are still reasonable. For instance, the model with $Q_p
= 10^{3.8}$, $e \sim$0.01, and a heavy-element core of $M_c = 30$
$M_{\oplus}$ produces the observed radius well, assuming
3$\times$solar atmospheric opacity.  To better constrain models of
TrES-4, definitive measurements of, or stronger limits to, its
orbital eccentricity are needed.

\subsection{XO-3b}

XO-3b has been observed to be supermassive and on an eccentric
orbit \citep[$M_p = 13.25_{-0.64}^{+0.64}$ $M_J$, $e =
0.260_{-0.017}^{+0.017}$;][]{johnskrull07}. The discoverers
obtained transit light curves with relatively small 0.3-m
telescopes, spectroscopic observations using the 2.7-m Harlan J.
Smith (HJS) telescope and the 11-m Hobby-Eberly Telescope (HET) ,
and RV measurements with the HJS telescope. Based on theoretical
spectral models of the HJS data, \citet{johnskrull07} derive
$T_{eff} = 6429_{-50}^{+50}$ K, [Fe/H]$_{\ast}$ =
$-0.177_{-0.027}^{+0.027}$ dex, and $\log g_{\ast} =
3.95_{-0.062}^{+0.062}$ for the star. Combined with the RV
measurements, they arrive at $M_{\ast} = 1.41_{-0.08}^{+0.08}$
$M_{\odot}$ and $R_{\ast} = 2.13_{-0.21}^{+0.21}$ $R_{\odot}$ for
the star, and $M_p = 13.25_{-0.64}^{+0.64}$ $M_J$ and $R_p =
1.95_{-0.16}^{+0.16}$ $R_J$ for the planet. These authors have
commented that XO-3b is observed to be so large that in all cases
analyzed by \citet{fortney07a}, their models predict a much
smaller radius. However, due to the absence of a precise
trigonometric parallax of XO-3, its distance is very uncertain.
Assuming a smaller distance, and, hence, a reduced stellar mass and
radius than obtained using the isochrone method,
\citet{johnskrull07} found a best fit to their transit light
curves with $\log g_{\ast} = 4.19$, $M_{\ast} = 1.24 M_{\odot}$,
and $R_{\ast} = 1.48$ $R_{\odot}$, with the corresponding
estimates for $M_p$ of $12.03_{-0.46}^{+0.46}$ $M_J$ and $R_p$ of
$1.25_{-0.15}^{+0.15}$ $R_J$. These light-curve-based results
have recently been strengthened by \citet{winn08}
using larger aperture telescopes.  These authors
observed 13 transits photometrically using the 1.2-m
telescope at the FLWO, along with 0.4-0.6-m telescopes. Based
on these more precise transit light curves, they concluded that
$\log g_{\ast} = 4.244$, $M_{\ast} = 1.213 M_{\odot}$, and
$R_{\ast} = 1.377$ $R_{\odot}$, with the corresponding estimates
for $M_p$ of $11.79_{-0.59}^{+0.59}$ $M_J$ and for $R_p$ of
$1.217_{-0.073}^{+0.073}$ $R_J$. Since a trigonometric parallax
measurement of XO-3 is still lacking that could distinguish
these two different methods, we make models for both the
spectroscopically-determined and light-curve-based parameter sets.

Both the spectroscopical results of \citet{johnskrull07} and the
light-curve-based results of \citet{winn08} are listed in Table
\ref{tab:planet}. We note that the $M_p$ and $R_p$ values derived
using the two different methods differ significantly from one
another ($\sim$10\% in $M_p$ and $\sim$50\% in $R_p$). Therefore,
separate models and discussions for these different sets of planetary
properties are in order and we calculate theoretical radii for
XO-3b for both estimates of the planetary radius and mass.

\subsubsection{Results for XO-3b}\label{sec:res:xo3}

\begin{figure}
 \centering
    \includegraphics[width=89mm]{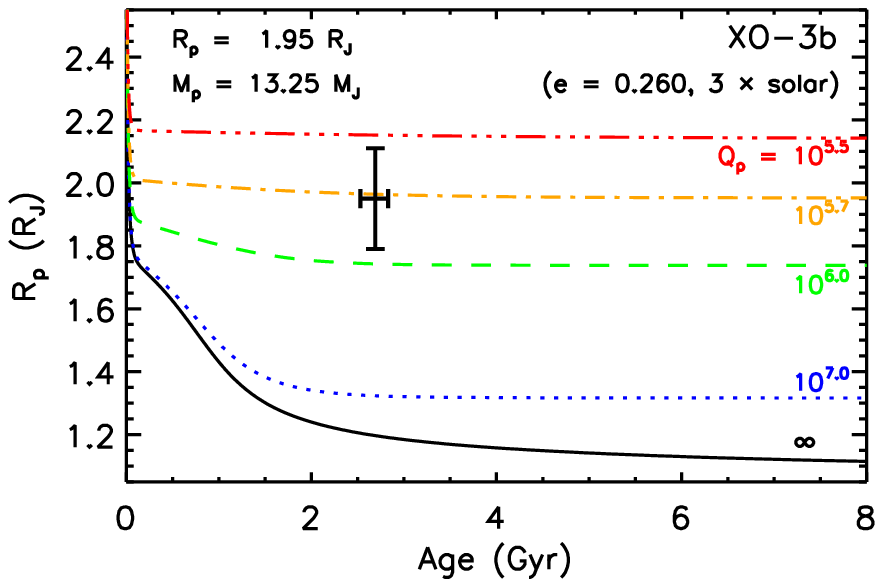}
    \includegraphics[width=89mm]{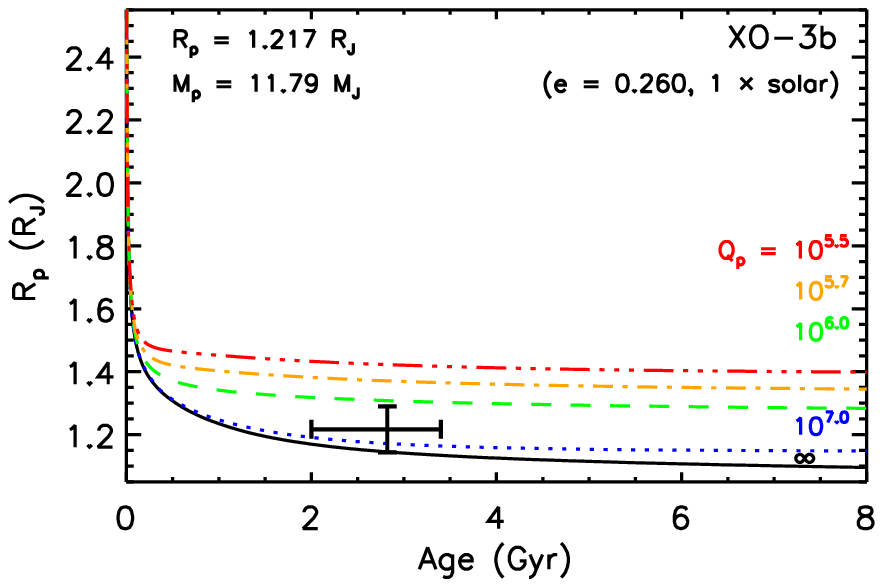}
   \caption{Theoretical planet radius $R_p$ ($R_J$) versus age (Gyr) for
   XO-3b. In the first panel, a planet mass of $M_p = 13.24$ $M_J$ based on the spectral
   synthesis model of \citet{johnskrull07} and 3$\times$solar atmospheric opacity are assumed,
   whereas in the second panel the corresponding values are $M_p = 11.79$ $M_J$ according
   to the light-curve fitting results of \citet{winn08} and 1$\times$solar atmospheric opacity. Also shown with error
   bars on both panels are the observed radii and age estimates from \citet{johnskrull07} and \citet{winn08},
   based on the two different analyses. In both panels, models assuming various tidal
   heating rates corresponding to
   $Q_p = \infty$ (no heating), 10$^{7.0}$, 10$^{6.0}$, 10$^{5.7}$, and 10$^{5.5}$ are color coded.
   In all the models, $e = 0.260$ \citep{johnskrull07} is assumed. Note that the observed radii, based on the
   two different methods, differ quite a bit from one another ($\sim$50\%). The very large radius based on
   spectral synthesis modeling can be fit by the model with $Q_p\sim$$10^{5.7}$, whereas the much
   smaller one inferred from the light-curve fit can be explained by models with $Q_p$ values down
   to $10^6$, within 1-$\sigma$ errors.  See \S\ref{sec:res:xo3} for more discussion.}
   \label{fig:xo3}
\end{figure}
%-------------

We include the possible heating due to
orbital tidal dissipation, assuming reasonable values of $Q_p$.
Our results for XO-3b are shown in Fig. \ref{fig:xo3}, where in
the first panel a planet mass of $M_p = 13.24$ $M_J$, derived from
the spectral synthesis method of \citet{johnskrull07}, and
an atmospheric opacity associated with 3$\times$solar metallicity are assumed, whereas in the
second panel the corresponding values are $M_p = 11.79$ $M_J$ from
the light-curve fitting results of \citet{winn08} and
1$\times$solar atmospheric opacity. In both cases, we assume that
a massive heavy-element core is absent. For such massive planets,
the effect of reasonable core masses on planetary radii is small.

As shown in the first panel of Fig. \ref{fig:xo3}, the very large
radius derived by \citet{johnskrull07}, based on spectral
synthesis modeling, can be explained by energy dissipation due to
tidal heating when the $Q_p$ parameter of XO-3b is
$\sim$$10^{5.7}$. For the case of the light-curve-based planetary
mass ({second panel} of Fig. \ref{fig:xo3}), the inferred radius
is broadly consistent to within 1-$\sigma$ errors with the absence
of an internal heat source, or with tidal heating with $Q_p
\gtrsim 10^6$. These specific constraints on $Q_p$ assume
3$\times$solar in the former case and 1$\times$solar atmospheric
opacity in the latter. For our first model of XO-3b assuming $M_p
= 13.24$ $M_J$ and 3$\times$solar atmospheric opacity, it would be
surprising to find a planet with $R_p/R_J \lesssim 1.2$ or
$\gtrsim 2.2$, since this will require $Q_p$ to be either too large
or too small. For the second model assuming $M_p = 11.79$ $M_J$
and 1$\times$solar atmospheric opacity, such a region would be
$R_p/R_J \lesssim 1.1$ or $\gtrsim 1.6$, for the same reason.
Given significant eccentricity, it is important to appropriately
account for tidal heating in order to model the planet's
structural evolution. We want to emphasize our adopted model
assumptions that 1) the planet is in synchronous rotation and that
2) all the tidal heating is in the convective core. Even though the
$e^2/Q_p$ degeneracy is broken due to the known value of $e$,
detailed radius evolution models could be used to constrain $Q_p$
for EGPs, but only if it is determined that tidal dissipation occurs
predominantly either in the convective core or in the atmosphere,
and if the uncertainties in the core mass and atmospheric opacity are
both resolved.

\subsection{HAT-P-1b}

Using photometry conducted by the Hungarian-made Automated
Telescope Network (HATNet) project, \citet{bakos07a} discovered
HAT-P-1b transiting one member of the stellar binary ADS 16402.
These authors suggested that HAT-P-1b was too large to be
explained by theoretical EGP models. Spectral synthesis modeling
of the parent star ADS 16402B, based on its Keck spectra, yielded
$T_{eff} = 5975_{-45}^{+45}$ K and [Fe/H]$_{\ast}$ =
$+0.13_{-0.02}^{+0.02}$ dex. \citet{bakos07a} also fit both
stellar members in the binary to evolutionary tracks and based on
the Subaru and the Keck spectra derived $M_{\ast} =
1.12_{-0.09}^{+0.09} M_{\odot}$ and $R_{\ast} =
1.15_{-0.07}^{+0.10} R_{\odot}$ for ADS 16402B and a best-fit age
of $3.6$ Gyr for the binary. Using the $z$-band transit curves
from KeplerCam \citep{holman06}, combined with RV measurements
from Subaru and Keck, these authors derive $M_p =
0.53_{-0.04}^{+0.04} M_J$ and $R_p = 1.36_{-0.09}^{+0.11} R_J$,
where the errors in the planetary radius include both statistical
and systematic errors in both the stellar radius and mass. Note
that in the above fits, a circular orbit ($e = 0$) was assumed.
However, the $\chi^2$ fitting of the RV data in \citet{bakos07a}
favors a small, yet non-zero, eccentricity:
$e=0.09_{-0.02}^{+0.02}$. These authors did estimate the heating
rate assuming $e = 0.09$ and suggested that if this non-zero
orbital eccentricity is confirmed, the observed large $R_p$ could
be explained by tidal heating. However, note that a non-zero
eccentricity is only suggestive, mainly due to the small number of
RV observations (13 velocities, for which the typical S/N is about
150 per pixel). Since RV-based eccentricity estimates are
positively biased due to noise \citep[e.g.,][]{shen08}, it is very
likely that the true $e$ is smaller than $0.09$. In fact,
\citet{johnson08} find a upper limit on $e$ of 0.067, with 99\%
confidence, by combining their new and previous RV measurements.
Therefore, we assume a smaller value, $e = 0.01$, in our baseline
model and see where such an assumption leads. Such a small
eccentricity could result from Kozai cycles with tidal friction
\citep[e.g.,][]{fabrycky07}, though there is evidence that the
spin-orbit misalignment is small \citep{johnson08}.

Based on more high-precision transit observations, however,
\citet{winn07d} report that HAT-P-1b is less ``bloated'' than
originally thought. Their observations include three transits
observed in $z$ band with the 1.2-m telescope at the FLWO, three
observed through the ``Gunn Z'' filter \citep{pinfield97} using
the Nickel 1-m telescope at Lick Observatory, and three observed
through the Johnson $I$ filter using the 1-m telescope at the Wise
Observatory. \citet{winn07d} derived $R_{\ast}/M_{\ast}^{1/3}$ by
fitting the transit light curves, and concluded that $R_{\ast} =
1.115_{-0.043}^{+0.043} R_{\odot}$ and $R_p =
1.203_{-0.051}^{+0.051} R_J$. Note that in their fits the orbital
eccentricity was assumed to be zero. These authors suggest that
the updated radius can be explained by the structural models of
\citet{burrows07b}, unless the planet has a very massive core of
heavy elements.  Indeed, \citet{burrows07b} calculated radius-age
trajectories for HAT-P-1b. They included different core masses and
atmospheric opacities in their models, and found that in order to
fit the observed radius, HAT-P-1b deviates from the
stellar-metallicity versus core-mass sequence otherwise roughly
followed by the transiting EGPs included in their paper \citep[see
Fig. 9 of][]{burrows07b}. However, the stellar and planetary
parameters of the HAT-P-1 system adopted by \citet{burrows07b} are
from the discovery work of \citet{bakos07a}. The parameters of the
HAT-P-1 system derived by \citet{torres08} are listed in Table
\ref{tab:planet}. These authors compile the $z$-band light curves
of \citet{winn07d} and the RV data and the atmospheric parameters
of \citet{bakos07a}, but with increased uncertainties for
$T_{eff}$, [Fe/H]$_{\ast}$, and $\log g_{\ast}$. As we show in
\S\ref{sec:res:hatp1}, using these new parameters and the new
planet radius, we now find inferred core masses that are roughly
consistent with the stellar-metallicity versus core-mass
relationship followed by the EGPs studied by \citet{burrows07b}.

%-------------
\begin{figure}
 \centering
    \includegraphics[width=89mm]{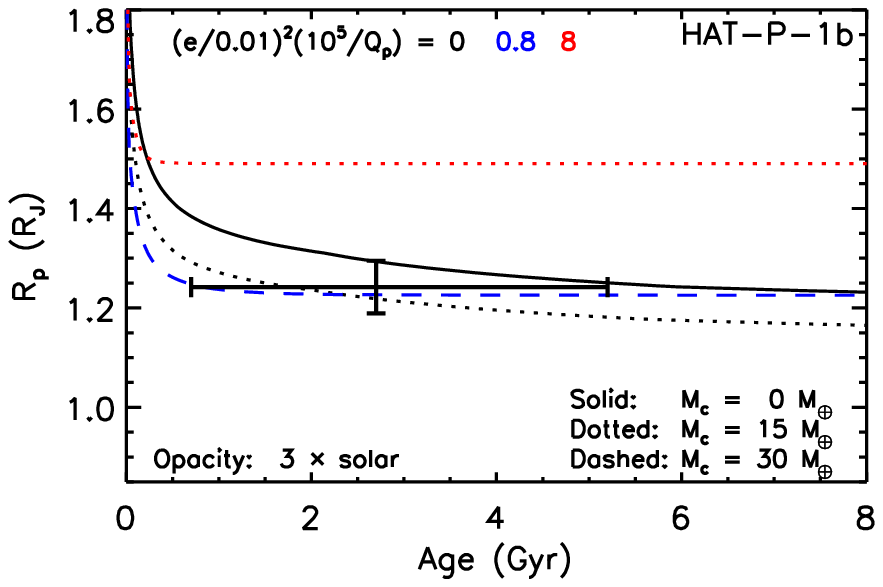}
    \includegraphics[width=89mm]{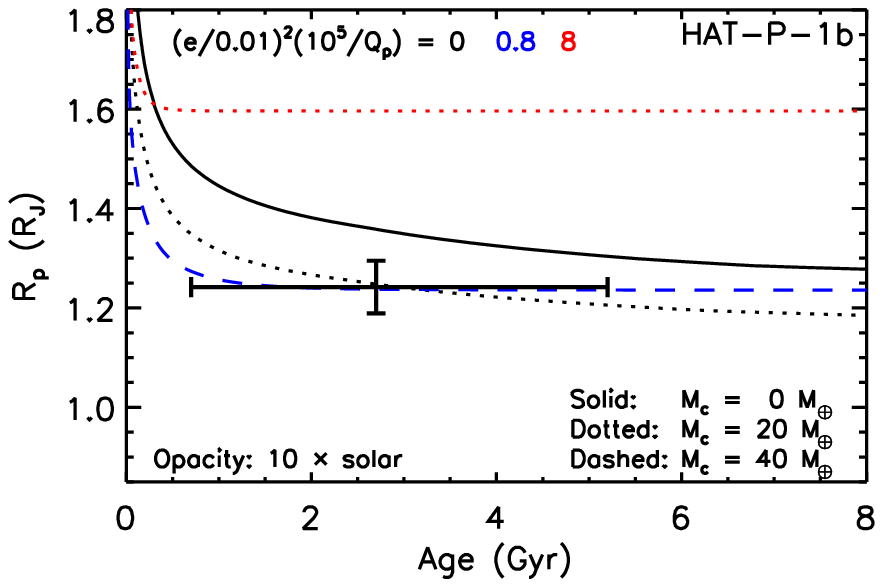}
   \caption{Theoretical planet radius $R_p$ ($R_J$) versus age (Gyr)
   for HAT-P-1b. Also shown with error bars are the observational stellar age and planet
   radius from \citet{torres08}, as listed in Table \ref{tab:planet}. We reexamine the best-fit
   core masses for HAT-P-1b, under various assumptions concerning the atmospheric opacity and the
   possible core heating rate due to tidal dissipation. The top (bottom) panel shows models assuming
   3$\times$solar (10$\times$solar) atmospheric opacities. In both panels, the black, blue, and
   red curves denote models assuming $(e/0.01)^2(10^5/Q_p) = $ 0, 0.8, and 8, respectively.
   Different line styles correspond to models with various heavy-element core masses in units
   of the Earth mass, $M_{\oplus}$, as labeled on the plot. If there is tidal heating assuming
   reasonable values of $(e/0.01)^2(10^5/Q_p) \sim$$0.8$, then the core mass required to fit
   the observed radius is $\sim$30 $M_{\oplus}$ ($\sim$40 $M_{\oplus}$), assuming 3$\times$solar
   (10$\times$solar) atmospheric opacity, in which case HAT-P-1b does follow the core-mass
   stellar-metallicity relation found by \citet{burrows07b}.  See \S \ref{sec:res:hatp1} for a
   discussion.}
   \label{fig:hatp1}
\end{figure}
%-------------

\subsubsection{Results for HAT-P-1b}\label{sec:res:hatp1}

Given the new measured radii and stellar age of \citet{torres08},
we have reexamined the best-fit core masses for HAT-P-1b. The
effects of the possible heating due to tidal dissipation on the
planet's structural evolution are considered, assuming a small yet
non-zero eccentricity $e = 0.01$. Theoretical evolutionary
trajectories of planet radius with age for HAT-P-1b are shown in
Fig. \ref{fig:hatp1}, where the first (second) panel displays the
results assuming a 3$\times$solar (10$\times$solar) atmospheric
opacity. Models with different tidal heating rates proportional to
$(e/0.01)^2(10^5/Q_p)$ are color-coded as labeled. For both of the
panels, different line styles represent ``no heavy-element
cores," or the presence of a heavy-element core with a range of
masses in units of Earth masses, $M_{\oplus}$. For clarity, only
selected models are presented in the figure. Table \ref{tab:hatp1}
lists the best estimates for the core mass under various
assumptions.

As demonstrated in Fig. \ref{fig:hatp1} and Table \ref{tab:hatp1},
there are multiple solutions to explain the observed radius of
HAT-P-1b. In all the cases considered, a non-zero heavy-element
core mass is needed, which, without any external heat sources, is
$\sim$15 $M_{\oplus}$ ($\sim$20 $M_{\oplus}$) assuming
3$\times$solar (10$\times$solar) atmospheric opacity,. This
best-fit core mass becomes larger when there is external heating.
If HAT-P-1b does follow the approximate
core-mass/stellar-metallicity relation found by
\citet{burrows07b}, then given its [Fe/H]$_{\ast}$
($+0.13_{-0.08}^{+0.08}$ dex), the core mass would be $\sim$30
$M_{\oplus}$ ($\sim$40 $M_{\oplus}$) assuming 3$\times$solar
(10$\times$solar) atmospheric opacity. These core masses
correspond to the cases with $(e/0.01)^2(10^5/Q_p) = 0.8$.

In summary, if there is tidal heating with reasonable values of
$Q_p$, the core-mass estimates suggest that HAT-P-1b follows the
correlation between stellar metallicity and core mass found by
\citet{burrows07b}, or if there is no extra heating, deviates
mildly from the correlation sequence. However, a larger core mass,
more in keeping with the correlation found by \citet{burrows07b},
is more consistent with reasonable values of
$(e/0.01)^2(10^5/Q_p)$, as long as $e$ is non-zero and $Q_p$ is
not anomalously small.

%------------------------------------------------------------------------------

\section{Equilibrium Planetary Radii and Equilibration Timescales for Various Heating Rates: a Parameter Study}\label{sec:heating}

%------------------------------------------------------------------------------

\begin{figure*}
 \centering
    \includegraphics[width=166mm]{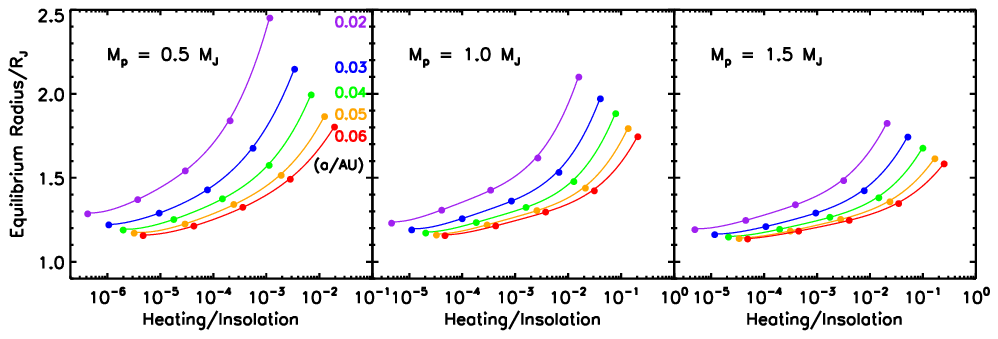}
    \includegraphics[width=166mm]{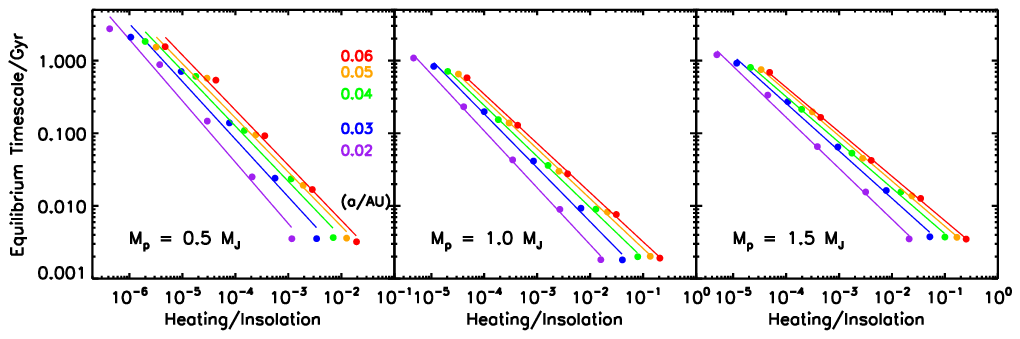}
   \caption{Equilibrium planetary radius $R_{eq}$ ($R_J$) and equilibration timescale
   (Gyr) assuming various ratios between the core heating
   power and the insolation power. The equilibration timescale is defined as the
   time it takes the planet to evolve from 1.25 $R_{eq}$ to 1.05 $R_{eq}$. Models are calculated
   for planets with masses $M_p =$ 0.5, 1.0, and 1.5 $M_J$, and semi-major axes $a =$ 0.02,
   0.03, 0.04, 0.05, and 0.06 AU, orbiting a G2V star. Filled circles represent results
   calculated from radius-age trajectories, whereas the curves are fits to them given by
   eq. (\ref{eq:fitting}) and the corresponding parameters in Table \ref{tab:fitting}.
   See \S \ref{sec:heating} for more information.}
   \label{fig:equiradiage}
\end{figure*}
%-------------

In this section, we investigate the effects of generic core
heating on the planet's equilibrium radius, and the time to reach
this equilibrium. We refer to the latter as the equilibration
timescale. The heat source discussed in this section could be due
to orbital tidal heating, but does not have to be. Figure
\ref{fig:equiradiage} shows equilibrium planetary radii ($R_{eq}$)
and equilibration timescales ($\tau_{eq}$) with various
core-heating powers, quantified by
$\dot{E}_{heating}/\dot{E}_{insolation}$. Models are calculated
for planets with masses $M_p =$ 0.5, 1.0, and 1.5 $M_J$, with
semi-major axes $a =$ 0.02, 0.03, 0.04, 0.05, and 0.06 AU,
orbiting a G2V star. Equilibrium is defined as the state after
which the planet radius is constant to within a part in $10^{5}$.
We define the equilibration timescale as the time it takes the
planet to evolve from $1.25$ $R_{eq}$ to 1.05 $R_{eq}$. Note that
for those rare models for which the planetary radii still change by more than
$10^{-5}$ $R_{eq}$ at the end of calculation (10 Gyr), the
equilibrium state is assumed to have been reached at the final age
of the evolutionary trajectory.

In Fig. \ref{fig:equiradiage}, filled circles represent the values
of $R_{eq}$ and $\tau_{eq}$ calculated from our theoretical
trajectories, whereas the curves are least-square fits to them. We
adopt a fourth-order polynomial in fitting $R_{eq}/R_J$ as a
function of log$(\dot{E}_{heating}/\dot{E}_{insolation})$, and a
linear model for log$(\tau_{eq}/{\rm Gyr})$ versus
log$(\dot{E}_{heating}/\dot{E}_{insolation})$, given by:
\begin{equation}\label{eq:fitting}
\begin{array}{rcl}
\displaystyle \frac{R_{eq}}{R_J} & = & \displaystyle C_0 + C_1 x + C_2 x^2 + C_3 x^3 + C_4 x^4, \\[3mm]
\displaystyle \log \bigg(\frac{\tau_{eq}}{{\rm Gyr}}\bigg) & = & \displaystyle b + k x,
\end{array}
\end{equation}
where $x \equiv \log(\dot{E}_{heating}/\dot{E}_{insolation})$. The
model fits of the parameters are given in Table \ref{tab:fitting}.

For an extreme close-in EGP with $M_p = 0.5$ $M_J$ at $a = 0.02$
AU orbiting a G2V star, the equilibrium planetary radii range from
$\sim$1.3 $R_J$ for little heating
($\dot{E}_{heating}/\dot{E}_{insolation} \lesssim 10^{-6}$) to
$\sim$2.5 $R_J$ for strong heating
($\dot{E}_{heating}/\dot{E}_{insolation} \sim 10^{-3}$). The
corresponding timescales for the planet to reach these radii are
$\sim$3 Gyr for the former and $\sim$3 Myr for the latter. At $a =
0.06$ AU, the equilibrium radii are smaller and the relevant
timescales are longer $-$ from $R_{eq} \sim 1.25$ $R_J$ and
$\tau_{eq}\sim 2$ Gyr for $\dot{E}_{heating}/\dot{E}_{insolation}
\lesssim 10^{-5}$ to $R_{eq} \sim 1.8$ $R_J$ and $\tau_{eq}\sim 3$
Myr for $\dot{E}_{heating}/\dot{E}_{insolation} \sim 10^{-2}$. For
more massive planets, the equilibrium radii are smaller and the
timescales are longer. For an EGP with $M_p = 1.5$ $M_J$ at $a =
0.02$ AU, the values are $R_{eq}\sim 1.8$ $R_J$ and $\tau_{eq}\sim
5$ Myr for $\dot{E}_{heating}/\dot{E}_{insolation} \sim 10^{-2}$.
Our theoretical model grids along with the fitting curves provided
in eq. (\ref{eq:fitting}) and the parameters listed in Table
\ref{tab:fitting} can be used to calculate the equilibrium
planetary radius and the typical timescale to reach it, given
different combinations of planet mass, orbital distance, and the
ratio of core-heating power to insolation power.

%------------------------------------------------------------------------------
\section{Summary and Discussion}\label{sec:sum}

We have calculated theoretical radius-age trajectories for three
EGPs: TrES-4, XO-3b, and HAT-P-1b, under various assumptions
concerning atmospheric opacity, the presence of an inner
heavy-element core, and possible heating due to orbital tidal
dissipation. The main model results are the following:
\begin{enumerate}

\item[1.] Unless the atmospheric opacity of TrES-4 is unusually
large (much higher than 10$\times$solar equivalent), core heating
is required to explain its observed radius \citep[$R_p =
1.751_{-0.062}^{+0.064}$ $R_J$;][]{torres08}. However, the
required heating power is modest. A non-core model with
$(e/0.01)^2(10^5/Q_p) = 0.1$ produces radii consistent with the
1-$\sigma$ lower boundary of $R_p$ from \citet{torres08}, assuming
10$\times$solar atmospheric opacity. The required energy
dissipation rates become larger for models with a heavy-element
core. The model with $Q_p = 10^{3.8}$, $e \sim$0.01, and a
heavy-element core of $M_c = 30$ $M_{\oplus}$, reproduces the
observed radius well, assuming 3$\times$solar atmospheric opacity.
If TrES-4 follows the core-mass stellar-metallicity correlation
found by \citet{burrows07b}, then the models with a non-zero
heavy-element core mass are favored, considering its stellar
metallicity [Fe/H]$_{\ast}$ = $+0.14_{-0.09}^{+0.09}$ dex
\citep{torres08}. Ongoing Spitzer photometry of its secondary
eclipse will put more stringent constraints on $e$ and can either
confirm or rule out these possibilities.

\item[2.] For XO-3b, we have shown that orbital tidal heating is a
key factor in explaining the planet radius. The very large radius
($R_p = 1.95_{-0.16}^{+0.16}$ $R_J$) derived by
\citet{johnskrull07} based on spectral synthesis modeling can be
explained by energy dissipation due to tidal heating. In this
case, the $Q_p$ parameter of XO-3b is near $\sim$$10^{5.7}$, a not
unreasnoable value. On the other hand, the much smaller radius
($1.217_{-0.073}^{+0.073}$ $R_J$) based on light-curve fit by
\citet{winn08} is consistent with no core heating sources, or with
tidal heating assuming $Q_p \gtrsim 10^6$, within 1-$\sigma$
errors. These constraints on $Q_p$ assume 3$\times$solar
atmospheric opacity for the former case and 1$\times$solar for the
latter, but are only weakly dependent on this.

\item[3.] We have reexamined the core mass required for HAT-P-1b
using the updated data (importantly, its radius) from
\citet{torres08}, and now find it generally follows the
correlation between core mass and stellar metallicity found by
\citet{burrows07b}. In all the cases considered, a non-zero
heavy-element core mass is needed to explain the observed radius
\citep[$R_p = 1.242_{-0.053}^{+0.053}$ $R_J$;][]{torres08}.  The
core mass is $\sim$15 $M_{\oplus}$ ($\sim$20 $M_{\oplus}$)
assuming 3$\times$solar (10$\times$solar) atmospheric opacity when
there is no external heating. If there is tidal heating
corresponding to reasonable values of $(e/0.01)^2(10^5/Q_p)
\sim$$0.8$, then the core mass required to fit the observed radius
is $\sim$30 $M_{\oplus}$ ($\sim$40 $M_{\oplus}$) assuming
3$\times$solar (10$\times$solar) atmospheric opacity, in which
case HAT-P-1b follows the core-mass stellar-metallicity relation
found by \citet{burrows07b} and \citet{guillot06}.

\end{enumerate}

In addition, we have carried out a parameter study of the effects
of core heating and provided theoretical grids and fitting
formulae for the equilibrium planet radius and equilibration
timescale, given various core heating powers for planets with
masses $M_p =$ 0.5, 1.0, and 1.5 $M_J$ with $a =$ 0.02-0.06 AU,
orbiting a G2V star. The fitting formula for the equilibrium
planet radius can be used for a theoretical zeroth-order estimate,
without carrying out detailed evolutionary calculations. The
equilibration timescale $\tau_{eq}$ characterizes the time it
takes the planet to adjust its structure in response to a given
degree of core heating.

Recently \citet{jackson08a} considered the effect of the
co-evolution of the orbital eccentricity and the semi-major axis
on the tidal dissipation history.  In the past, the semi-major
axis had been assumed to be constant when conducting tidal
evolution studies
\citep[e.g.][]{bodenheimer01,bodenheimer03,gu04}.
\citet{jackson08b} calculate the evolutionary histories of the
tidal dissipation rate for several EGPs, and find that in most
cases the tidal heating rate increases as a planet moves inward
and then decreases as the orbit circularizes. The relevant
timescale, $\tau_{heating}$, is the time it takes the tidal
heating rate to decay by a factor of $e$. If $\tau_{eq} \gg
\tau_{heating}$, then it is valid to take a constant effective
tidal heating rate in the planet radius-age trajectory
calculation. If $\tau_{eq} \sim$$\tau_{heating}$, then in order to
account for the effect of a varying tidal heating rate, different
values should be used at each time step of the radius-age
trajectory calculation. If $\tau_{eq} \ll \tau_{heating}$, then
the planet will have enough time to reach an equilibrium radius
before the tidal heating rate decays significantly. In this case,
the planet's structure evolves in a quasi-equilibrium manner.
Theoretical planet radius-age trajectory models will easily be able to
account for the effect of varying tidal heating rate by
adopting different core heating rates at each time interval of
$\tau_{heating}$ during the calculation. In effect, there is a
reset of the ``clock'' right after the time of maximum heating $-$
the planet becomes most extended on a timescale $\sim$$\tau_{eq}$
after the tidal heating rate achieves this maximum. Because of the
intense heating and the quick response, the planet loses the
memory of its shrinkage history before maximum heating, which
is effectively a reset of its ``age.''

\citet{burrows07b} calculated theoretical radii for HD 209458b.
They found that the measured radius deviated at the
$\sim$1.5-$\sigma$ level for the age they assumed, even when
employing 10$\times$solar atmospheric opacity, no inner solid
core, and no core heating. However, the updated age measurement
for HD 209458b by \citet{torres08} of $3.1_{-0.7}^{+0.8}$ Gyr is
much smaller than the one adopted by \citet{burrows07b}
($5.5_{-1.5}^{+1.5}$ Gyr), whereas the updated radius,
$1.359_{-0.019}^{+0.016}$, is similar to that used by
\citet{burrows07b} ($1.32_{-0.03}^{+0.03}$). As a result, the
\citet{torres08} radius and age measurement for HD 209458b can be
fit by the 10$\times$solar-opacity model of \citet{burrows07b}
within $\sim$1-$\sigma$ uncertainties, without the need of any
core heating sources. Moreover, if the tidal heating rate of HD
209458b decayed from $\sim$4$\times10^{26}$ erg s$^{-1}$ to
$\sim$4$\times10^{24}$ erg s$^{-1}$ during the past 2 Gyr
\citep{jackson08b}, then its effective age would be 2 Gyr younger,
due to the ``clock reset'' effect.  Based on our parameter study
in \S\ref{sec:heating}, the equilibration timescale of HD 209458b
during maximum heating would have been $\sim$0.05 Gyr. This is
small enough compared with the decay timescale of the tidal
heating rate ($\sim$1 Gyr from $\sim$4$\times10^{26}$ erg s$^{-1}$
to $\sim$1.5$\times10^{26}$ erg s$^{-1}$) for our models to
reproduce HD 209458b's current radius.  In this case, the
\citet{torres08} radius measurement for HD 209458b, along with an
``effective age'' of $\sim$1.1 Gyr, can explain HD 209458b's
radius within $\sim$1-$\sigma$ uncertainties, even with the
1$\times$solar-opacity model of \citet{burrows07b} and a core mass
of 10-20 $M_{\oplus}$, but without the need for any {current} core
heating. The latter comports with the very small limit of
$\sim$0.001 on its current orbital eccentricity. A possible caveat to the
tidal evolution scenarios described in \citet{jackson08a,jackson08b} is
that they can be dramatically changed due to even a small undetected
perturbing body \citep[e.g.,][]{mardling07}.  Sensitive searches
for such companion bodies are needed to further constrain this
possibility. Nevertheless, future studies should combine orbital semi-major
axis, eccentricity, and planet radius evolution models in a more
coupled fashion. It is not only more consistent to consider the
co-evolution of $a$, $e$, and $R_p$, but also important to factor
in the feedback of the associated tidal heating power on $R_p$ and
its radius-age trajectory \citep{jackson08a,jackson08b}. Such a
project is in progress.

%------------------------------------------------------------------------------
\acknowledgments

We thank G\'{a}sp\'{a}r Bakos for comments on the eccentricity of
HAT-P-1b, and Georgi Mandushev for insights concerning the
possible eccentricity range of TrES-4. We also thank Josh Winn for
a careful reading of an earlier version of the manuscript, and an
anonymous referee for a careful and useful report that improves
the paper. This study was supported in part by NASA grants
NNG04GL22G, NNX07AG80G, and NNG05GG05G and through the NASA
Astrobiology Institute under Cooperative Agreement No.
CAN-02-OSS-02 issued through the Office of Space Science.

%----------------------------------------------------------------------------

%-------------------------------------------------------------

\begin{appendix}

\section{External Heating due to Tidal Dissipation}

The total tidal energy dissipation rate within the planet in its
rest frame assuming equilibrium tides with constant lag angle and
synchronous rotation is
\citep[e.g.][]{goldreich66,bodenheimer01,bodenheimer03,gu04}:
\begin{equation}
\begin{array}{rcl}
\displaystyle \dot{E}_{{\rm tide}} & = & \frac{GM_{\ast}\mu f(e)}{a\tau_{{\rm
circ}}} \\[3mm]
& \approx & \displaystyle 1.1 \times 10^{24}\, {\rm erg}\,{\rm s}^{-1}
\bigg(\frac{e}{0.01}\bigg)^2\bigg[\frac{f(e)}{e^2}\bigg]\,
\bigg(\frac{M_{\ast}}{M_{\odot}}\bigg)\bigg(\frac{M_p}{M_J}\bigg)\bigg(\frac{a}{0.05
\, {\rm AU}}\bigg)^{-1}\bigg(\frac{\tau_{{\rm circ}}}{{\rm
Gyr}}\bigg)^{-1} \,
%&\approx 1.1 \times 10^{25}\,
%\bigg(\frac{e}{0.01}\bigg)^2\bigg[\frac{f(e)}{e^2}\bigg]\,\bigg(\frac{Q_p}{10^5}\bigg)^{-1}
%\bigg(\frac{M_{\ast}}{M_{\odot}}\bigg)^{5/2}\bigg(\frac{R_p}{R_J}\bigg)^{5}\bigg(\frac{a}{0.05
%\, {\rm AU}}\bigg)^{-15/2} \, ,
\end{array}
\end{equation}
where $\mu\equiv M_{\ast}M_p/(M_{\ast} + M_p)$ is the reduced
mass, $f(e) \equiv \frac{2}{7}[h_3(e) - 2h_4(e) + h_5(e)]$ is a
function of orbital eccentricity with
$h_3(e)=(1+3e^2+3e^4/8)(1-e^2)^{-9/2}$,
$h_4(e)=(1+15e^2/2+45e^4/8+5e^6/16)(1-e^2)^{-6}$, and
$h_5(e)=(1+31e^2/2+255e^4/8+185e^6/16+25e^8/64)(1-e^2)^{-15/2}$
\citep{gu04}. Note that $f(e) \rightarrow e^2$ as $e \rightarrow 0$.
$\tau_{{\rm circ}}$ denotes the circularization timescale, which
is
\begin{equation}
%\begin{split}
\tau_{{\rm circ}} \approx 0.10 \,{\rm Gyr} \times
\bigg(\frac{Q_p}{10^5}\bigg)\bigg(\frac{M_{\ast}}{M_{\odot}}\bigg)^{-3/2}\bigg(\frac{M_p}{M_J}\bigg)\bigg(\frac{R_p}{R_J}\bigg)^{-5}\bigg(\frac{a}{0.05
\, {\rm AU}}\bigg)^{13/2}\,  .
%\end{split}
\end{equation}

A more informative quantity is the ratio of the tidal energy
dissipation rate and the insolation rate, given by
{\footnotesize
\begin{equation}
\begin{array}{rcl}
\displaystyle \frac{\dot{E}_{{\rm tide}}}{\dot{E}_{{\rm insolation}}} & = &
\displaystyle \frac{GM_{\ast}\mu f(e)}{\pi F_p R_p^2 \,a\tau_{{\rm circ}}} \\[3mm]
& \approx & \displaystyle 6.9 \times 10^{-5}
\bigg(\frac{e}{0.01}\bigg)^2\bigg[\frac{f(e)}{e^2}\bigg]\,\bigg(\frac{Q_p}{10^5}\bigg)^{-1}
\bigg(\frac{M_{\ast}}{M_{\odot}}\bigg)^{5/2}\bigg(\frac{R_p}{R_J}\bigg)^{3}\bigg(\frac{a}{0.05
\, {\rm AU}}\bigg)^{-15/2}\bigg(\frac{F_p}{10^9 \,{\rm erg}\,{\rm
cm}^{-2}\,{\rm s}^{-1}}\bigg)^{-1}.\\
\end{array}
\end{equation}
}

\end{appendix}

\bibliographystyle{apj}
\bibliography{apj-jour,epa}

\clearpage
%----------------------------------------------------------------------------------------------
\begin{landscape}
\begin{deluxetable}{cccccccccccccc}
\tabletypesize{\scriptsize} \tablewidth{0pc} \tablecaption{
Observational Properties of the Transiting Planet Systems.
\label{tab:planet} } \tablehead{ \colhead{} & \colhead{$a$} &
\colhead{Period} & \colhead{} & \colhead{$M_*$} & \colhead{$R_*$}
& \colhead{$T_{eff}$} & \colhead{[Fe/H]$_*$} & \colhead{Age} &
\colhead{$M_p$} & \colhead{$R_p$} &
\colhead{$F_p$\tablenotemark{a}} & \colhead{$\dot{E}_{{\rm
tide}}/\dot{E}_{{\rm insolation}}$\tablenotemark{b}} &
\colhead{} \\
\colhead{~~~~~System~~~~~} & \colhead{(AU)} & \colhead{(days)} &
\colhead{$e$} & \colhead{($M_{\odot}$)} & \colhead{($R_{\odot}$)}
& \colhead{(K)} & \colhead{(dex)} & \colhead{(Gyr)} &
\colhead{($M_J$)} & \colhead{($R_J$)} & \colhead{($10^9$ erg
cm$^{-2}$ s$^{-1}$)} & \colhead{[$(Q_p/10^5)^{-1}$]} &
\colhead{References} } \startdata
 XO-3\dotfill    & $0.0476_{-0.0005}^{+0.0005}$    & 3.19154    & $0.260_{-0.017}^{+0.017}$    & $1.41_{-0.08}^{+0.08}$    & $2.13_{-0.21}^{+0.21}$    & $6429_{-50}^{+50}$     & $-0.177_{-0.027}^{+0.027}$  & $2.69_{-0.16}^{+0.14}$ & $13.25_{-0.64}^{+0.64}$       & $1.95_{-0.16}^{+0.16}$     & $4.20_{-0.84}^{+0.84}$ & $5.9\times 10^{-1}$                   & \tablenotemark{c}  \\ [+1.2ex]
     \dotfill    & $0.0454_{-0.00082}^{+0.00082}$  & 3.19152    & $0.260_{-0.017}^{+0.017}$    & $1.213_{-0.066}^{+0.066}$ & $1.377_{-0.083}^{+0.083}$ & $6429_{-100}^{+100}$   & $-0.177_{-0.080}^{+0.080}$  & $2.82_{-0.82}^{+0.58}$ & $11.79_{-0.59}^{+0.59}$       & $1.217_{-0.073}^{+0.073}$  & $1.93_{-0.27}^{+0.27}$ & $3.1\times 10^{-1}$                   & \tablenotemark{d}  \\ [+1.2ex]
 TrES-4\dotfill  & $0.05092_{-0.00069}^{+0.00072}$ & 3.553945   & $\sim 0.0$                   & $1.394_{-0.056}^{+0.060}$ & $1.816_{-0.062}^{+0.065}$ & $6200_{-75}^{+75}$     & $+0.14_{-0.09}^{+0.09}$     & $2.9_{-0.4}^{+0.4}$    & $0.920_{-0.072}^{+0.073}$     & $1.751_{-0.062}^{+0.064}$  & $2.31_{-0.20}^{+0.21}$ & $2.3\times 10^{-4}$ \tablenotemark{e} & \tablenotemark{f}  \\ [+1.2ex]
 HAT-P-1\dotfill & $0.0553_{-0.0013}^{+0.0012}$    & 4.46543    & $\sim 0.0$                   & $1.133_{-0.079}^{+0.075}$ & $1.135_{-0.048}^{+0.048}$ & $5975_{-120}^{+120}$   & $+0.13_{-0.08}^{+0.08}$     & $2.7_{-2.0}^{+2.5}$    & $0.532_{-0.030}^{+0.030}$     & $1.242_{-0.053}^{+0.053}$  & $0.66_{-0.08}^{+0.08}$ & $9.2\times 10^{-5}$ \tablenotemark{e} & \tablenotemark{f}  \\ [+1.2ex]
\enddata
\tablenotetext{a}{The stellar flux at the planet's substellar
point.} \tablenotetext{b}{$\dot{E}_{{\rm tide}}$ is the total
tidal energy dissipation rate within the planet in its rest frame
(Gu et al. 2004). $\dot{E}_{{\rm insolation}} \equiv \pi
R_p^2F_p$.} \tablenotetext{c}{\citet{johnskrull07}; Inferred from
spectroscopically derived stellar parameters.}
\tablenotetext{d}{\citet{winn08}; Determined from light-curve
fits.} \tablenotetext{e}{Assuming $e = 0.01$.}
\tablenotetext{f}{\citet{torres08}; Note that they assume $e = 0$
exactly in deriving the parameters, since the radial-velocity data
are consistent with a circular orbit.}
\end{deluxetable}
\clearpage
\end{landscape}
%-------------

%----------------------------------------------------------------------------------------------

\begin{deluxetable}{cccccccc}
\tabletypesize{\scriptsize} \tablecaption{Best-estimate Core
Masses for HAT-P-1b under Various Assumptions. \label{tab:hatp1} }
\tablewidth{0pt} \tablehead{ \colhead{Atmospheric Opacity} &
\multicolumn{3}{c}{$3 \times$solar} & \colhead{} &
\multicolumn{3}{c}{$10 \times$solar} \\
\cline{2-4}
\cline{6-8} \\
\colhead{$\bigg(\frac{e}{0.01}\bigg)^2\bigg(\frac{10^5}{Q_p}\bigg)$}
& \colhead{0} & \colhead{0.8} & \colhead{8} & \colhead{} &
\colhead{0} & \colhead{0.8} & \colhead{8} } \startdata
$M_c/M_{\oplus}$\dotfill & 15 &  30 & 60  &  & 20  &  40 &  80 \\
\enddata
%%%%%%%
\tablecomments{The tidal dissipation rate is proportional to
$e^2/Q_p$ when $e \ll 1$.}
\end{deluxetable}

%----------------------------------------------------------------------------------------------
\begin{deluxetable}{ccccccccc}
\tabletypesize{\scriptsize} \tablewidth{0pc} \tablecaption{
Fitting Parameters for the Equilibrium Planetary Radius and
Equilibration Timescale for EGPs Undergoing Core Heating.
\label{tab:fitting} } \tablehead{ \colhead{$M_p$} & \colhead{$a$}
& \colhead{} & \colhead{} & \colhead{} & \colhead{} & \colhead{} &
\colhead{} &
\colhead{} \\
\colhead{($M_J$)} & \colhead{(AU)} &
\colhead{$C_0$\tablenotemark{a}} &
\colhead{$C_1$\tablenotemark{a}} &
\colhead{$C_2$\tablenotemark{a}} &
\colhead{$C_3$\tablenotemark{a}} &
\colhead{$C_4$\tablenotemark{a}} & \colhead{$b$\tablenotemark{b}}
& \colhead{$k$\tablenotemark{b}} } \startdata
 $0.5$      & $0.02$    & $15.4$    & $9.69$    & $2.63$    & $0.328$   & $0.0156$  & $-4.78$   & $-0.845$  \\ [+1.2ex]
 \dotfill   & $0.03$    & $7.74$    & $4.43$    & $1.22$    & $0.157$   & $0.00785$ & $-4.26$   & $-0.794$  \\ [+1.2ex]
 \dotfill   & $0.04$    & $6.42$    & $4.00$    & $1.25$    & $0.182$   & $0.0101$  & $-3.95$   & $-0.762$  \\ [+1.2ex]
 \dotfill   & $0.05$    & $4.61$    & $2.66$    & $0.866$   & $0.135$   & $0.00809$ & $-3.77$   & $-0.746$  \\ [+1.2ex]
 \dotfill   & $0.06$    & $3.60$    & $1.79$    & $0.573$   & $0.0905$  & $0.00560$ & $-3.71$   & $-0.758$  \\ [+1.2ex]
\tableline
 $1.0$      & $0.02$    & $6.42$    & $4.51$    & $1.58$    & $0.255$   & $0.0155$  & $-4.10$   & $-0.783$  \\ [+1.2ex]
 \dotfill   & $0.03$    & $4.44$    & $3.02$    & $1.16$    & $0.208$   & $0.0139$  & $-3.70$   & $-0.742$  \\ [+1.2ex]
 \dotfill   & $0.04$    & $3.42$    & $2.19$    & $0.894$   & $0.170$   & $0.0122$  & $-3.42$   & $-0.701$  \\ [+1.2ex]
 \dotfill   & $0.05$    & $2.72$    & $1.54$    & $0.669$   & $0.137$   & $0.0106$  & $-3.27$   & $-0.686$  \\ [+1.2ex]
 \dotfill   & $0.06$    & $2.33$    & $1.15$    & $0.512$   & $0.110$   & $0.00897$ & $-3.17$   & $-0.677$  \\ [+1.2ex]
\tableline
 $1.5$      & $0.02$    & $4.20$    & $2.54$    & $0.888$   & $0.145$   & $0.00899$ & $-3.59$   & $-0.702$  \\ [+1.2ex]
 \dotfill   & $0.03$    & $3.15$    & $1.76$    & $0.663$   & $0.118$   & $0.00796$ & $-3.21$   & $-0.654$  \\ [+1.2ex]
 \dotfill   & $0.04$    & $2.57$    & $1.34$    & $0.543$   & $0.105$   & $0.00762$ & $-3.01$   & $-0.630$  \\ [+1.2ex]
 \dotfill   & $0.05$    & $2.13$    & $0.924$   & $0.391$   & $0.0808$  & $0.00634$ & $-2.90$   & $-0.622$  \\ [+1.2ex]
 \dotfill   & $0.06$    & $1.91$    & $0.693$   & $0.292$   & $0.0610$  & $0.00486$ & $-2.82$   & $-0.612$  \\ [+1.2ex]
\enddata
\tablenotetext{a}{$R_{eq}/R_J = C_0 + C_1 x + C_2 x^2 + C_3 x^3 +
C_4 x^4$, where $x \equiv
\log(\dot{E}_{heating}/\dot{E}_{insolation})$.}
\tablenotetext{b}{$\log(\tau_{eq}/{\rm Gyr}) = b + k x$.}
\end{deluxetable}
%-------------

%-------------
%-------------

\end{document}